\documentclass[pra,showpacs,floatfix]{revtex4-1}
\usepackage{comment}
\usepackage{xcolor}
\usepackage{graphics,graphicx,epsfig}
\usepackage[caption=false]{subfig}
\usepackage{amssymb,amsmath}
\usepackage{epstopdf}
\usepackage{amsmath}
\usepackage{mathtools}

\begin{document}

\title{Entanglement as a sufficient condition for quantum synchronization between two mechanical oscillators}
\author{Manju}
\email{2018phz0009@iitrpr.ac.in}
\author{Shubhrangshu Dasgupta, and Asoka Biswas}%
\affiliation{%
Department of Physics, Indian Institute of Technology Ropar, Rupnagar, Punjab 140001, India
}%

\date{\today}
\begin{abstract}
We present an optomechanical  model to show that entanglement can be a sufficient condition for quantum synchronization of two mechanical oscillators. As both these entities can be characterized in terms of variances of a set of EPR-like conjugate quadratures, we investigate whether this leads to a specific condition for simultaneous existence of the both. In our model, one of the oscillators makes the cavity, while the other is kept suspended inside the cavity, and the always-on coupling between the two is mediated via the same cavity mode. We show that in presence of amplitude modulation with the same frequency as that of the oscillators, these oscillators get nearly complete quantum synchronized and entangled simultaneously in the steady state. We also show  that entanglement always becomes accompanied by quantum synchronization, though the reverse is not necessarily true. Thus, entanglement becomes a sufficient condition for the quantum synchronization. This behaviour can be observed for a large range of system parameters. 
\end{abstract}
\maketitle
\section{INTRODUCTION}
Spontaneous synchronization is a natural phenomenon that can be often experienced in avian flight and flashing of fireflies \cite{strogatz}. It was also observed by Huygens in a classical clock pendulum  with a common support \cite{huygens1897oeuvres}. It has been explored in many different areas, namely, neuron networks \cite{leone2015synchronization,tang2011synchronization,cao2017fixed}, chemical reaction \cite{vaidyanathan2015dynamics}, nonlinear dynamics \cite{lee2013quantum1,walter2014quantum,shirasaka2017optimizing}, and electrical circuits and communications \cite{van1927frequency,pecora1990synchronization}. 

Classical synchronization between two nonlinearly coupled oscillators occurs due to a rephasing and energy redistribution of their motions. This refers to a limit cycle in phase space, where their respective generalized positions $q_j(t)$  and generalized linear momenta $p_j(t)$ ($j\in 1,2$) become equal.  In quantum regime, however, the one cannot measure these quadrataures with certainty at the same time $t$, according to Heisenberg's uncertainty principle. The best estimate of the quantum synchronization (QS) therefore corresponds to a state of two oscillators, for which the total uncertainty of their joint quadratures becomes minimum.  In this regard, Mari {\it et al.} proposed the following figure of merit \cite{mari2013measures}:
\begin{eqnarray}
S_{qm} = \left\langle q_{-}^{2}(t)+ p_{-}^{2}(t)\right\rangle^{-1}  \label{sqm}
\end{eqnarray}
in terms of the synchronization errors $ q_{-}(t) = \frac{1}{\sqrt{2}}\left[ q_{1}(t)- q_{2}(t)\right]$ and $ p_{-}(t) = \frac{1}{\sqrt{2}}\left[ p_{1}(t)- p_{2}(t)\right]$, where $\langle\cdots\rangle$ denotes the expectation (or mean) values of the corresponding operators. 
Expanding these operators around their mean values, 
$q_{-}(t)=Q_{-}(t)+\delta q_{-}(t)$, $ p_{-}(t)=P_{-}(t)+\delta p_{-}(t)$, and using the limit of zero mean of these quadrature differences, a modified form of the measure of QS can be obtained as  
\begin{equation}
S_{q}=\frac{1}{\left\langle \delta q_{-}(t)^{2}+\delta p_{-}(t)^{2}\right\rangle}\le 1\;.\nonumber\\
\label{SQ}
\end{equation}

On the other hand, quantum correlation between the interacting systems can also manifest itself as entanglement. 
In the context of two oscillators (two bosonic modes), entanglement has been characterized in terms of uncertainties of the joint quadratures. The criterion for entanglement can be derived starting from the uncertainty relation of $q_-$ and $p_-$ and then applying the Peres's separability criterion \cite{peres1996separability} based on the partial transposition. Any bimodal state can be said to be entangled if it violates the partially transposed uncertainty relation, as given by \cite{duan2000inseparability}
\begin{equation}
\langle(\delta q_-)^2\rangle+\langle (\delta p_+)^2\rangle \geq 1\;, \label{dc}
\end{equation}
where the partial transposition on the second subsystem changes $p_-$ into $p_+ = \frac{1}{\sqrt{2}}\left[ p_{1}+ p_{2}\right]$. A stronger version of the above inequality, as follows, can be obtained by using a simple algebraic identity \cite{mancini2002entangling,biswas2005NJP}:
\begin{equation}
E_D= \langle (\delta q_-)^2\rangle\langle (\delta p_+)^2\rangle\ge \frac{1}{4}\;.
\label{entcrit}
\end{equation}

It can be observed from the above, both the QS measure and the entanglement criterion are derived from the Heisenberg uncertainty principle for a pair of EPR-like variables. This suggests that both these seemingly different features originate purely from the same class of quantum correlation (namely, the EPR-type) and therefore can appear in the same system at a similar time-scale. In fact,  in our recent work \cite{garg2022quantum}, we have shown that it is indeed possible to both quantum-synchronize and entangle two oscillators in a certain parameter regime, by coupling them with two different cavities. In this paper, we will investigate if such possibility is generic in nature. With an always-on nonlinear coupling between two oscillators, we will show that they remain quantum-synchronized as long as entangled. More importantly, without the entanglement, they loose synchronization. 
Thus, the entanglement can be considered as a sufficient condition for achieving QS.

We note that whether two coupled systems can be simultaneously synchronized and entangled has been investigated in several systems. Manzano {\it et al.\/} showed that any two coupled oscillators in a network, initially prepared in a separable state, can be both synchronized and entangled  \cite{manzano2013synchronization}, for a suitable choice of the interaction strength to the other oscillators in the network. They further showed that in presence of synchronization, the entanglement is retained despite the decoherence. It was further shown in \cite{witthaut2017classical,lee2013quantum} that many-body system can be entangled in presence of synchronization. In the classical (quantum) limit, two single-mode cavities get synchronized (entangled). However, all the above works dealt only with classical synchronization. On the contrary, it is more meaningful to focus on QS, when one looks for its relationship with entanglement, which has no classical analogy. In this paper, we specifically investigate if there is any inherent quantum correlation that can lead to both QS and entanglement. Previous attempts in this regard include study of quantum discord \cite{giorgi2012,manzano2013synchronization}, linear entropy \cite{chimera},  mutual information based on von Neumann entropy \cite{ameri2015mutual} and Renyi entropy \cite{chimera2015}. In these works, similar steady state behavior of these measures of quantum correlation and the QS was reported. But their numerical studies could not essentially relate them as arising from the same physical origin. The present work addresses this exact loophole of the earlier reports. 

To do the relevant analysis, optomechanical system poses as a suitable platform. In our model, two mechanical oscillators indirectly interact with each other, via their common coupling to the same cavity mode. One of the  oscillators is suspended inside the cavity, while the other is usual movable mirror of the same cavity. Clearly, their interaction is always on  and weakly nonlinear, but still leads to near-complete QS and entanglement between them. We will specifically show that the quantum correlation manifests itself in the simultaneous existence of the two. This happens at the modulation frequency of the driving field. It is recently reported that the continuous measurement of a single bosonic system can enhance its phase synchronization with the external drive \cite{kwek}, when it is subjected to squeezed driving field, and both negative and nonlinear damping. A feedback control of the system based on the continuous measurement of an additional bath can also enhance synchronization \cite{nakao}. In this paper, we however consider a coupled system (not a single system entrained with a drive) and our result suggests an opposite effect upon measurement. If one of the oscillators is measured, it destroys the entanglement and hence the complete QS (note that we do not study the phase synchronization in this paper). In addition, the dynamics of the oscillators does not involve any two-photon processes, either driving or damping, unlike the van der Pol oscillators considered in \cite{kwek,nakao}.


We organize the paper in the following way. In the following Section, we introduce the model and the Hamiltonian, and derive all the equations of motion for the fluctuation operators. We obtain the analytic expressions of the fluctuations and its spectrum in Sec. III. In Sec. IV, we numerically solve the dynamics of the system and will study how both entanglement and QS between the oscillators can arise at similar time-scale. Finally, we conclude the paper in Sec. V.
\begin{figure}[h]
	\centering
	\includegraphics[width=.7\textwidth]{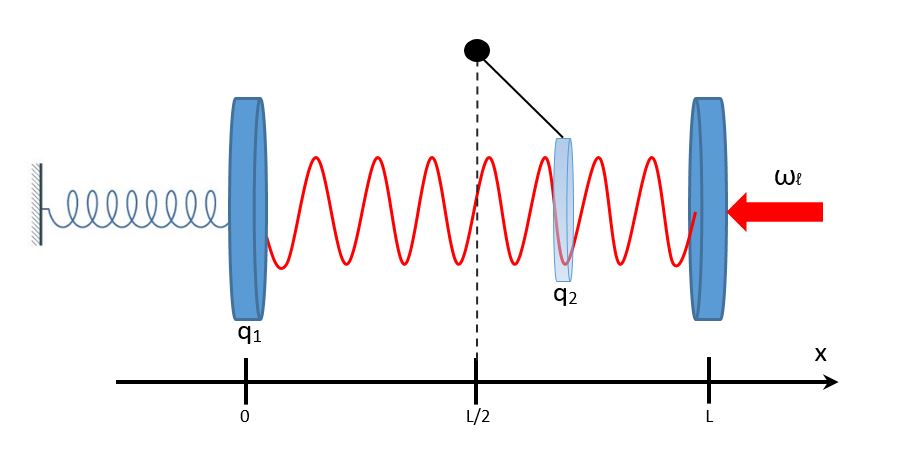}
	\caption{Schematic diagram of a membrane-in-middle optomechanical system (MIMOS), in which a membrane is suspended in the middle of a driven cavity.}
	\label{Model}
\end{figure}
\section{THEORETICAL MODEL}
The optomechanical system under consideration is shown in Fig. \ref{Model}. This system consists of an optical cavity with one mirror fixed, while the other is movable, acting as a mechanical oscillator (with resonance frequency $\omega_{m1}$). Another oscillating membrane is also suspended inside the cavity. The cavity is driven by a laser with frequency $\omega_{l}$ and time-modulated amplitude $E(t)$. This model is quite different from the other optomechanical setups in which two membranes (or mirrors) are optically coupled to two different cavities, while an additional mechanical coupling between them has to be externally introduced to facilitate the synchronization. In our model, the coupling between the mirror and the membrane is always on and there is no need to introduce any additional coupling. Their interaction with the same cavity mode generates an effective photon-number-dependent mirror-membrane coupling, which leads to their synchronization and entanglement as well. 

The coupling between the optical cavity and membrane depends upon the position of membrane, placed relative to the nodes and antinodes of cavity mode. An one-dimensional calculation gives the frequency of the cavity as a function of the mirror displacement $q_1$ and the membrane displacement $q_2$,  as follows \cite{thompson2008strong}:
\begin{eqnarray} 
\centering
   \omega_{cav}(q_{1},{q_{2}})=\left(\frac{c}{L+q_{1}}\right)\cos^{-1}\left(r_{c}\cos{\frac{4 \pi q_{2}}{\lambda}}\right)\;.
   \label{F1}
 \end{eqnarray}
 Here $L$ is the length of cavity, $r_{c}$ is the reflectivity of the membrane and $\lambda$ is the wavelength of laser field. Expanding $\omega_{cav}(q_{1},q_{2})$ about the equilibrium positions $q_{10}=0$ and $q_{20}$ of the respective cavity, up to second order in Taylor series, we have:
\begin{equation} 
\centering
   \omega_{cav}(q_{1},q_{2}) = \omega_{c} - g_{1}^{(1)} q_{1}-g_{1}^{(2)} q_{2}+ g_{2}^{(1)} q_{1}^{2}+ g_{2}^{(2)} q_{2}^{2}- g_{3} q_{1}q_{2}\;, \label{omegac}
\end{equation} 
where we can identify
$$\omega_c=\omega_{cav}(0,q_{20})-\left.\frac{\partial\omega_{cav}}{\partial q_2}\right|_{(0,q_{20})}q_{20}+\frac{1}{2} \left.\frac{\partial^2\omega_{cav}}{\partial q_2^2}\right|_{(0,q_{20})}q_{20}^2\;,$$
and 
\begin{eqnarray}
g_{1}^{(1)}&=&-\left.\frac{\partial\omega_{cav}}{\partial q_1}\right|_{(0,q_{20})}+ \frac{1}{2}\left.\frac{\partial^2\omega_{cav}}{\partial q_1\partial q_2}\right|_{(0,q_{20})}q_{20}\;,\nonumber\\
g_{1}^{(2)}&=&-\left.\frac{\partial\omega_{cav}}{\partial q_2}\right|_{(0,q_{20})}+ \left.\frac{\partial^2\omega_{cav}}{\partial q_2^2}\right|_{(0,q_{20})}q_{20}\;,\nonumber\\
g_{2}^{(j)}&=&\frac{1}{2} \left.\frac{\partial^2\omega_{cav}}{\partial q_j^2}\right|_{(0,q_{20})} \;\;,\;\; j = 1,2\nonumber\\
g_3&=&-\frac{1}{2} \left.\frac{\partial^2\omega_{cav}}{\partial q_1\partial q_2}\right|_{(0,q_{20})}\;.
\end{eqnarray}
The Hamiltonian of the system then takes the following form ($\hbar=1$):
\begin{equation}\label{hamil}
 H  =  \omega_{cav}(q_{1},q_{2}) a^{\dagger}a + \sum_{j=1,2} \frac{\omega_{mj}}{2} \left (q_{j}^{2}+p_{j}^{2}\right )
 + \iota E \left[1+\eta_{D}\cos(\Omega_{D}t)\right]\left(a^{\dagger}-a\right)\;,
\end{equation}    
where $\omega_{mj}$ $q_{j}$, and $p_{j}$ are  the frequency, dimensionless position  and momentum operators, respectively,  of the $j$th oscillator, $a^{\dagger} (a)$ is the creation (annihilation) operator of cavity mode, satisfying the commutation relation $\left[a,a^{\dagger}\right]=1$, and $\eta_{D}$ and $\Omega_{D}$ are the amplitude and the frequency of the modulating field. The quadratures satisfy the following commutation relation:  $\left[q_{j},p_{j'}\right] = \iota \delta_{jj'}.$

On putting the expression (\ref{omegac}) of $\omega_{cav}(q_{1},q_{2})$, the Hamiltonian in the frame rotating with the laser frequency $\omega_{l}$ can be written as  
\begin{eqnarray}
H  =  \Delta a^{\dagger}a + \sum_{j=1,2} \left[\frac{\omega_{mj}}{2} \left (q_{j}^{2}+p_{j}^{2}\right)+ \left(-g_{1}^{(j)}q_{j}+g_{2}^{(j)}q_{j}^{2}\right)a^{\dagger}a\right]-g_{3}q_{1}q_{2}a^{\dagger}a+\iota E \left[1+\eta_{D}\cos(\Omega_{D}t)\right]\left(a^{\dagger}-a\right)\;. \label{finalH}
\end{eqnarray}
Here $\Delta = \omega_{c}-\omega_{l}$ denotes the cavity mode detuning.
Considering the dissipation of the system, we can now obtain the following quantum Langevin equations for the relevant operators
\begin{eqnarray}
\frac{d q_{j}}{d t} &=&\omega_{mj} p_{j}\;,\nonumber\\
\frac{d p_{j}}{d t}&=&-\omega_{mj} q_{j}+g_{1}^{(j)} a^{\dagger} a-2 g_{2}^{(j)} a^{\dagger} a q_{j}+g_{3} a^{\dagger} a q_{3-j}-\gamma_{mj} p_{j}+\xi_{j}(t)\;,\nonumber\\
\frac{d a }{d t}&=&-(\kappa+\iota \Delta) a+\iota\sum_{j=1,2}\left( g_{1}^{(j)} q_{j}-g_{2}^{(j)} q_{j}^{2}\right)a+\iota g_{3} q_{1}q_{2}a
+E\left[1+\eta_{D} \cos\left(\Omega_{D}t\right)\right]+\sqrt{2 \kappa} a_{i n}\;, 
\label{p1}
\end{eqnarray}
where we have used the Hamiltonian (\ref{finalH}).
Here $\kappa$ is the decay rate of cavity mode, $\gamma_{mj}$ is the damping rate of $j$th mechanical oscillator, and $a_{in}$ is the input noise operator with the following two-time correlation functions: \cite{walls2008input}:
 \begin{equation}
 \left\langle a_{in}(t) a_{in}^{\dag}(t')\right\rangle =  \delta(t-t')\;,\;\left\langle\ a_{in}^{\dag }(t) a_{in}(t')\right\rangle =    0 \;.\end{equation} 
 On the other hand, the thermal bath at an equilibrium temperature $T$ is described by a Brownian noise $\xi_j(t)$ with zero mean and the following two-time correlation function:
\begin{equation} 
\left\langle\xi_j(t)\xi_j\left(t^{\prime}\right)\right\rangle=\frac{\gamma_{mj}}{2 \pi \omega_{mj}} \int \omega e^{-\iota \omega\left(t-t^{\prime}\right)}\left[1+\operatorname{coth}\left(\frac{\hbar \omega}{2 k_{B} T}\right)\right] d \omega\;,
 \label{G1}
 \end{equation}
 where $k_{B}$ is the Boltzmann constant. For the case of large quality factor of the mechanical oscillator, the expression above reduces to the following Markovian-approximated form:
 \cite{giovannetti2001phase}:
 \begin{eqnarray}
  \left\langle\xi_{j}(t) \xi_{j}\left(t^{\prime}\right)\right\rangle=\gamma_{mj}\left( 2 \bar n_{ mj }+1\right) \delta\left(t-t^{\prime}\right)\;,
  \label{g2}
 \end{eqnarray}
 where $\bar n_{mj}=1 /\left[\exp \left(\hbar \omega_{mj} / k_{B} T\right)-1\right]$
 is the mean phonon number of the $j$th mechanical oscillator.  
 
 \subsection{Solution in mean-field approximation}
Since it is difficult to solve equation (\ref{p1}) analytically, we use the mean-field approximation to simplify the calculation by rewriting the  operators as the sum of their mean values and quantum fluctuation near mean value, i.e., $a\rightarrow A + \delta a$, $q_j\rightarrow { Q_j} + \delta q_j$, and $p_j\rightarrow {P_j} + \delta p_j$, where $A=\left\langle a \right\rangle$, $Q_j=\left\langle q_j \right\rangle$. and $P_j=\left\langle p_j \right\rangle$. Thus, the quantum Langevin equations can be split into two sets of equations, namely, (i) the nonlinear equations for the mean values:
\begin{eqnarray}
\frac{d Q_{j}}{d t} &=& \omega_{mj} P_{j}\;, \nonumber\\
\frac{d P_{j}}{d t} &=& -\omega_{mj} Q_{j}+\left(g_{1}^{(j)}+g_{3} Q_{3-j}\right) |\alpha|^2-2 g_{2}^{(j)} Q_{j}|\alpha|^2 -\gamma_{mj} P_{j}\;, \nonumber\\
\frac{d A}{d t} &=&-(\kappa+\iota \Delta)A+\iota\sum_{j=1,2}\left(g_{1}^{(j)}Q_{j}- g_{2}^{(j)}Q_{j}^{2}\right)A+\iota g_{3}Q_{1}Q_{2} A+E\left[1+\eta_{D} \cos\left(\Omega_{D}t\right)\right]\;,
\label{f1}
\end{eqnarray}
and (ii) the linearized equations for the quantum fluctuations:
 \begin{eqnarray}
 \frac{d}{d t} \delta q_{j} &=& \omega_{j} \delta p_{j}\;,\nonumber\\
 \frac{d}{d t} \delta p_{j} &=&-\left(\omega_{mj}+2g_{2}^{(j)}|A|^{2}\right) \delta q_{j}+g_{3} |A|^{2}\delta q_{3-j}+G_{j}\left(A\delta a^{\dagger}+A^{*} \delta a\right) -\gamma_{mj} \delta p_{j}+\xi_{j}(t)\;,\nonumber\\
 \frac{d}{d t} \delta a &= &-\iota F \delta a + \iota G_{1} A \delta q_{1}+\iota G_{2} A \delta q_{2}-\kappa \delta a+\sqrt{2 \kappa} a_{in}\;,\label{eq:f2}
 \end{eqnarray}   
where \begin{eqnarray}G_{j}&=&g_{1}^{(j)}-2g_{2}^{(j)}Q_{j}+g_{3}Q_{3-j}\;,\nonumber\\
F&=&\Delta-g_{1}^{(1)}Q_{1}-g_{1}^{(2)}Q_{2}+g_{2}^{(1)}Q_{1}^{2}+g_{2}^{(2)}Q_{2}^{2}-g_{3}Q_{1}Q_{2}\;.
\end{eqnarray}
Here, we have ignored the second and higher-order terms in fluctuations. 

In order to calculate the desired markers, $S_q$ and $E_D$, respectively for QS and entanglement, we need to obtain the solutions for the quadrature fluctuations of the oscillators. Solving the Eqs. (\ref{eq:f2}) will be convenient if we replace the intra-cavity field and the input noise operators by their quadratures, as well:  $\delta x=\frac{1}{\sqrt{2}}\left(\delta a^{\dagger}+\delta a\right)$, $\delta y=\frac{\iota}{\sqrt{2}}\left(\delta a^{\dagger}-\delta a\right)$, $\delta x_{in}=\frac{1}{\sqrt{2}}\left(\delta a^{\dagger}_{in}+\delta a_{in}\right)$, and $\delta y_{in}=\frac{\iota}{\sqrt{2}}\left(\delta a^{\dagger}_{in}-\delta a_{in}\right)$. Therefore, the Eqs. (\ref{eq:f2})  takes a simpler form, as given by
\begin{eqnarray}
\dot{M}(t)=BM(t)+J(t)\;,
\end{eqnarray}
where $M(t)^T=\left( \delta q_{1},\delta p_{1},\delta q_{2},\delta p_{2},\delta x,\delta y\right) $, the vector $J(t)$ as given below contains the noise terms:
\begin{eqnarray}
J(t)^T &=&\bigg(0,\xi_1,0,\xi_2,
\sqrt{2\kappa}\delta x_{in},\sqrt{2\kappa}\delta y_{in}\bigg)\;,
\end{eqnarray}
and 
\begin{eqnarray}B=
 \begin{pmatrix}
   0 & \omega_{m1} & 0 & 0 & 0 & 0\\
-\omega_{m1}-2g_{2}^{(1)}|A|^{2} & -\gamma_{m1} & g_{3}|A|^{2}& 0 & \sqrt{2}G_{1}\operatorname{Re}\left(A\right) & \sqrt{2}G_{1}\operatorname{Im} \left(A\right)\\
0 & 0 & 0 & \omega_{m2} & 0 & 0\\
g_{3}|A|^{2} & 0 & -\omega_{m2}-2g_{2}^{(2)}|A|^{2} & -\gamma_{m2} & \sqrt{2}G_{2} \operatorname{Re}\left(A\right) & \sqrt{2}G_{2} \operatorname{Im} \left(A\right)\\
-\sqrt{2}G_{1} \operatorname{Im}\left(A\right) & 0 & -\sqrt{2}G_{2} \operatorname{Im}\left(A\right) & 0 & -\kappa & F\\
\sqrt{2}G_{1} \operatorname{Re}\left(A\right)& 0 & \sqrt{2}G_{2} \operatorname{Re}\left(A\right) & 0 & -F & -\kappa\\
\end{pmatrix}
\label{Mat}
\end{eqnarray}
is a $6 \times 6$ time-dependent coefficient matrix.
 \begin{figure}
	\subfloat[\label{pq}]{%
		\includegraphics[height=4.5cm,width=.4\linewidth]{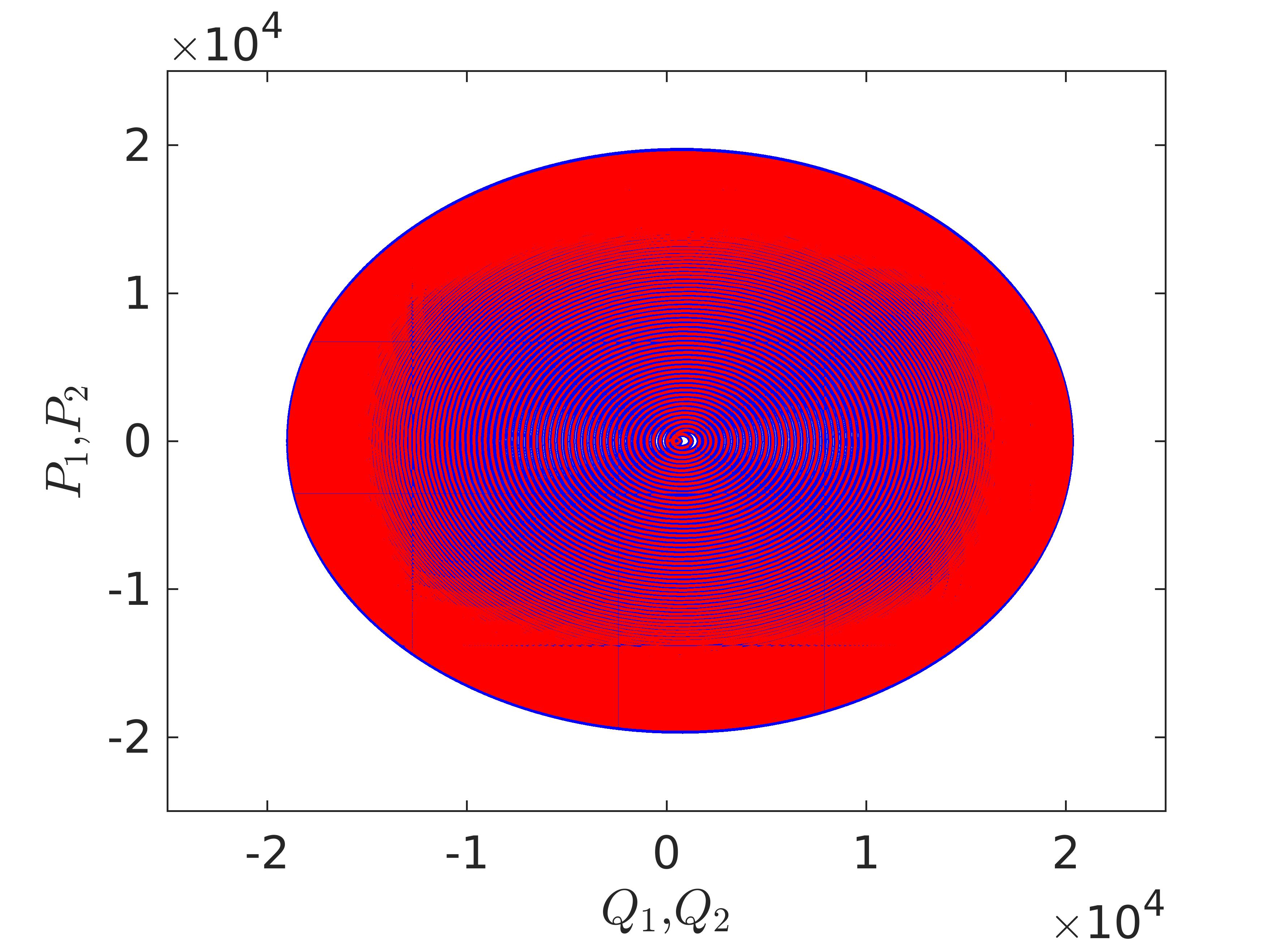}%
	}
	\subfloat[\label{q1q2}]{%
		\includegraphics[height=4.5cm,width=.4\linewidth]{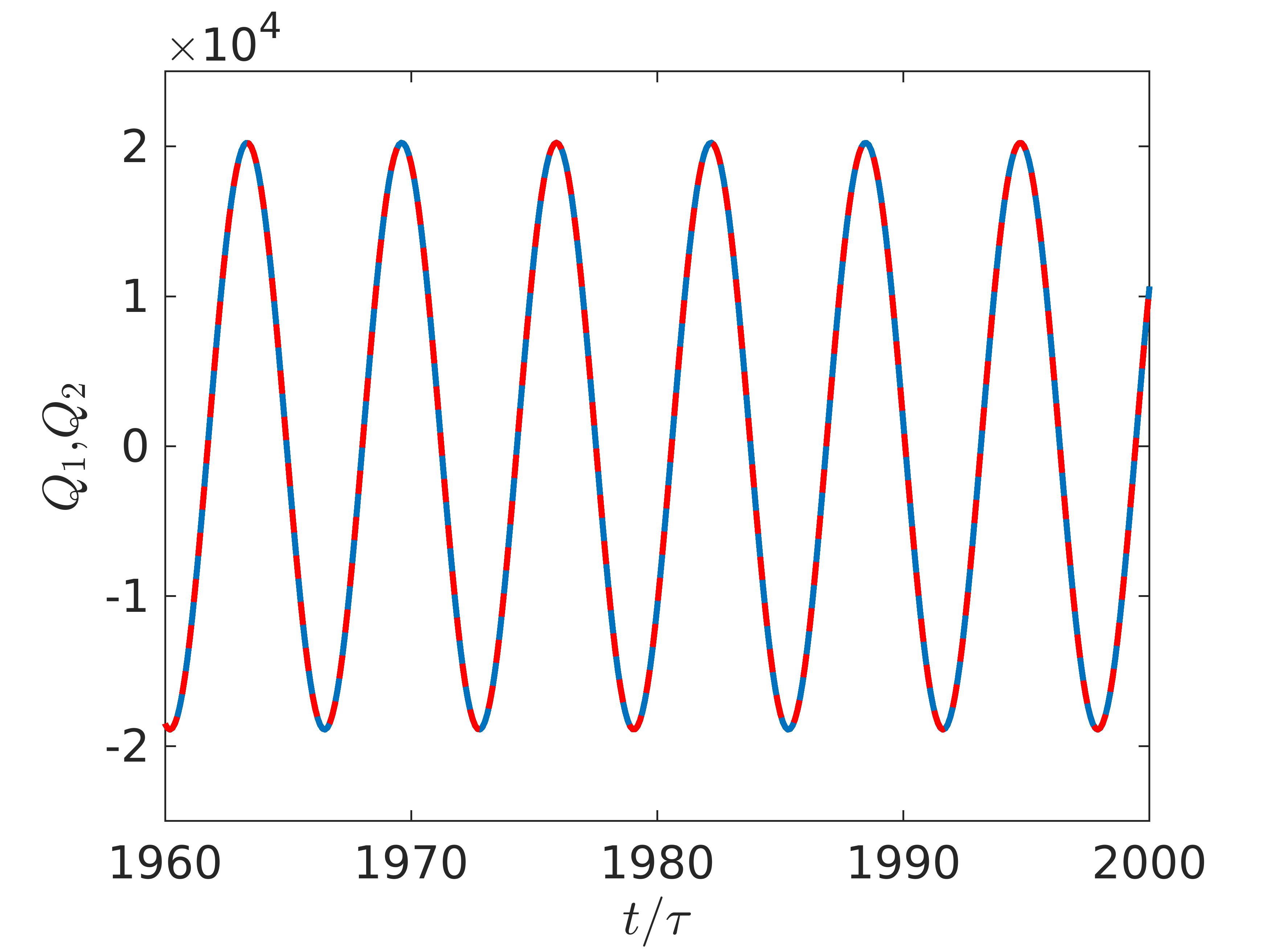}%
	}\\
	\subfloat[\label{p1p2}]{%
	    \includegraphics[height=4.5cm,width=.4\linewidth]{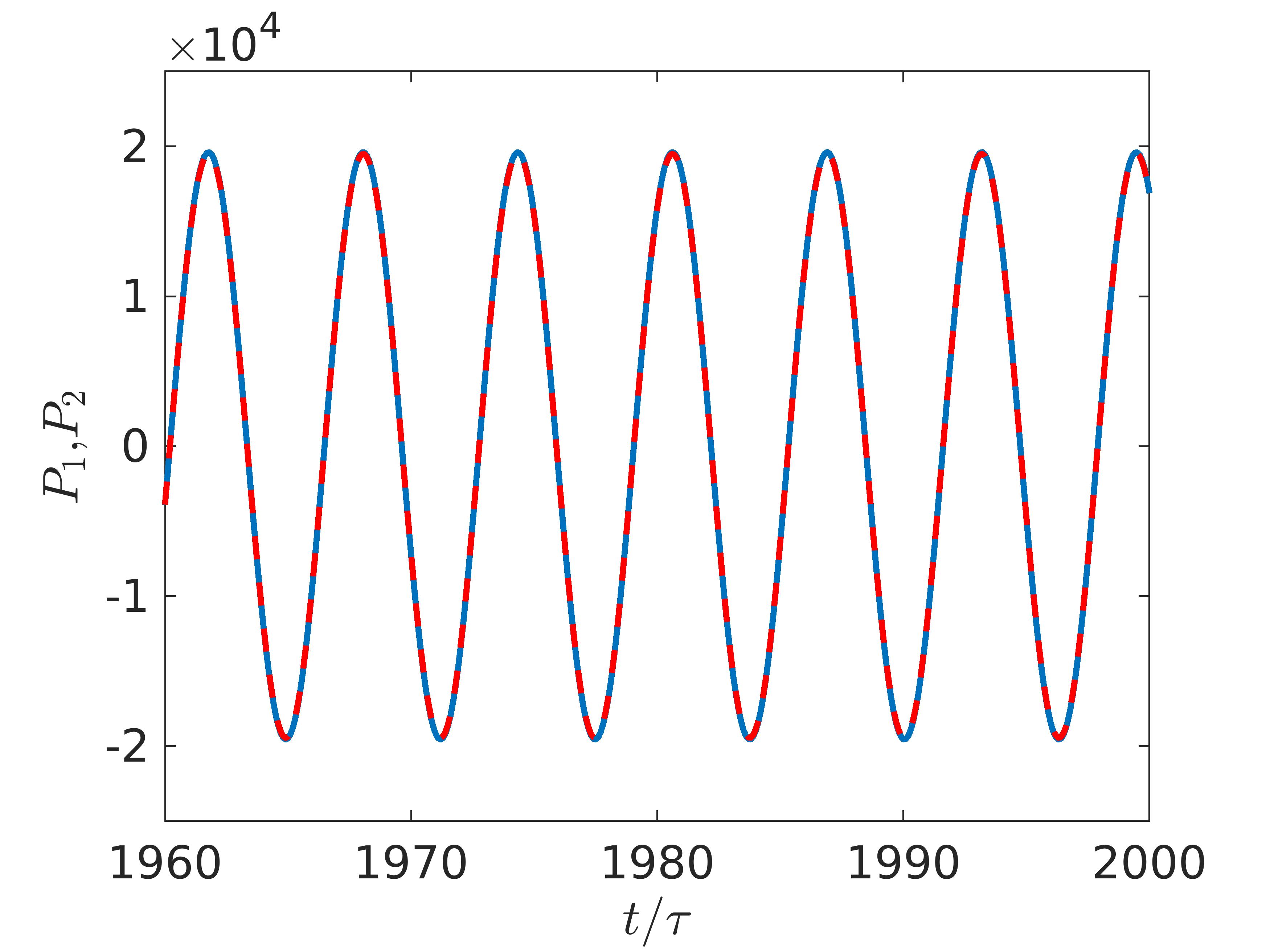}%
        }
        \subfloat[\label{Sq}]{%
	    \includegraphics[height=4.5cm,width=.4\linewidth]{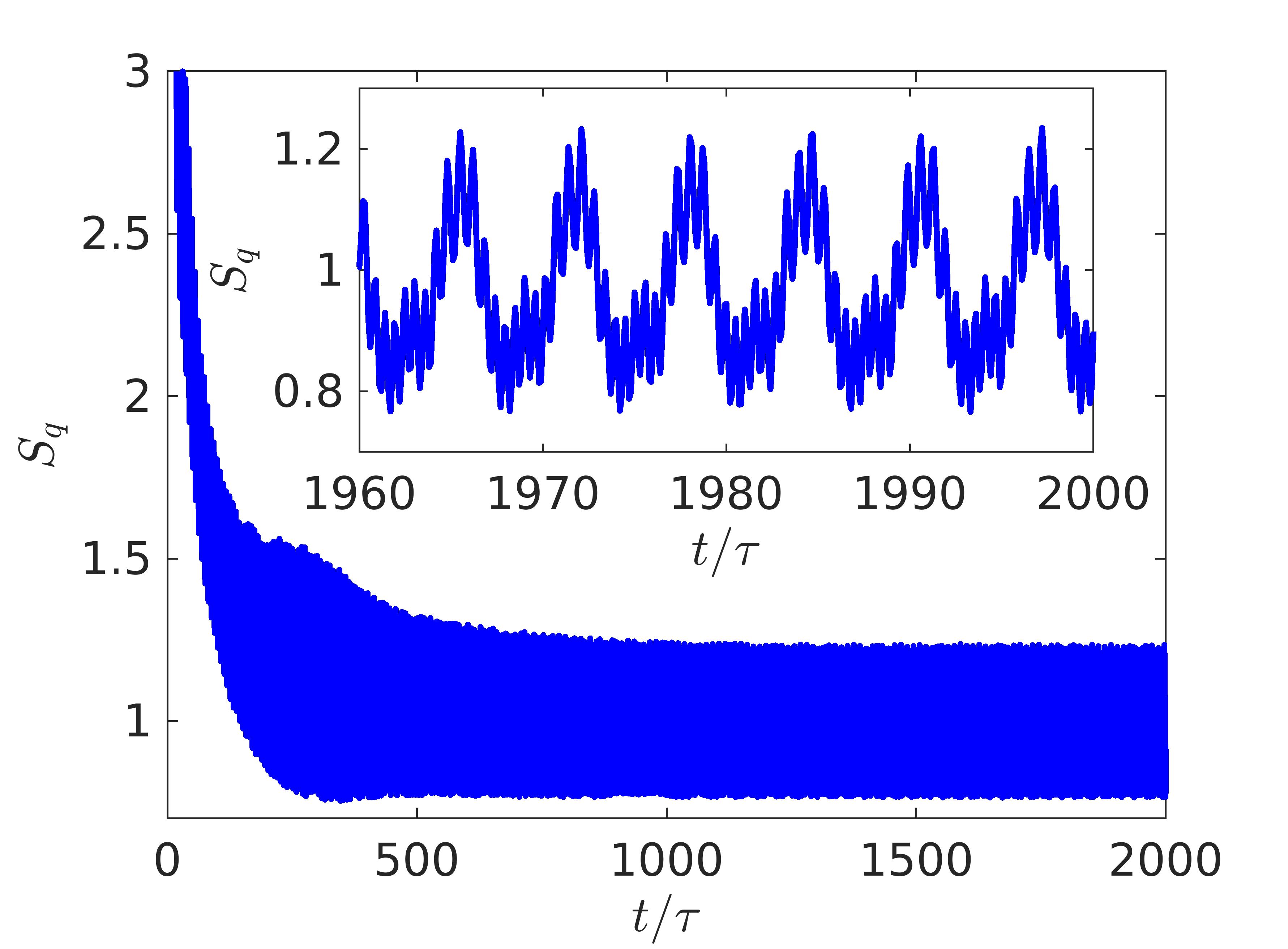}%
        }\\
        \subfloat[\label{ED}]{%
            \includegraphics[height=4.5cm,width=.4\linewidth]{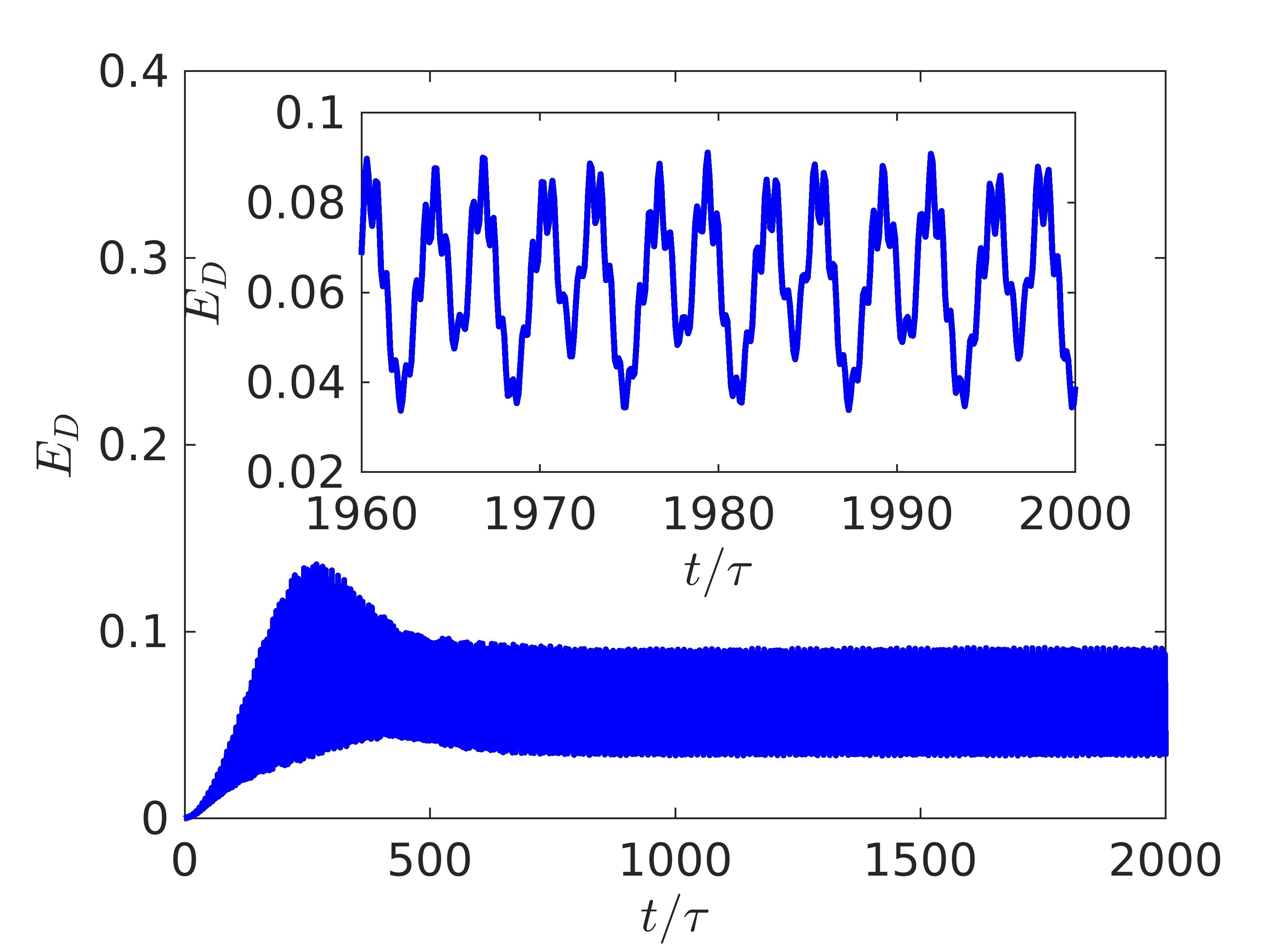}%
        }
       \subfloat[\label{delta}]{%
		\includegraphics[height=4.5cm,width=.4\linewidth]{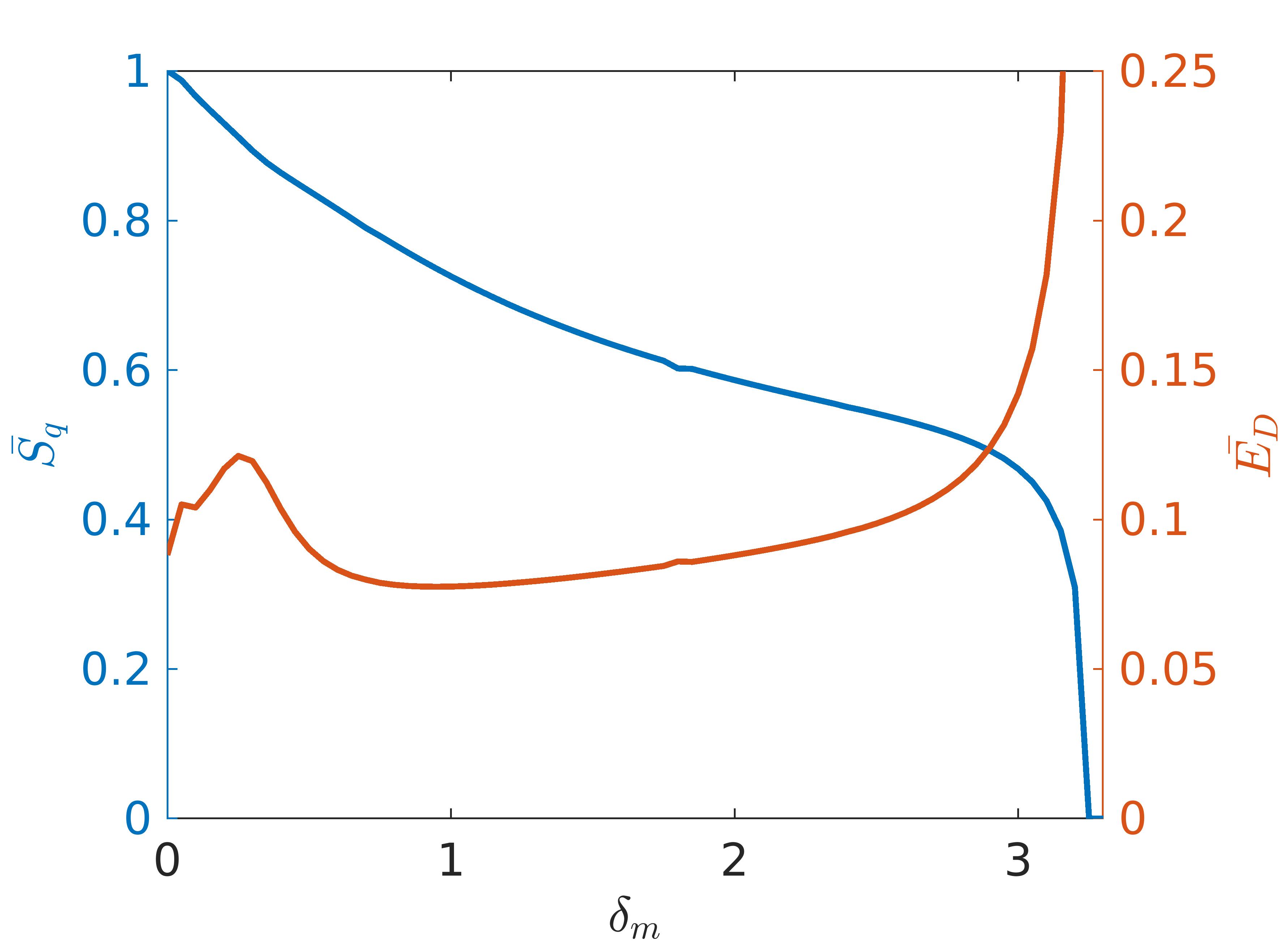}%
	}
\caption{(a) Limit-cycle trajectories in the $Q_{1}\leftrightharpoons P_{1}$ (red) and $Q_{2} \leftrightharpoons P_{2}$ (blue) spaces, Variation of (b) the mean values ${Q_1}$ (red) and ${Q_2}$ (blue), (c) the mean values ${P_1}$ (red) and ${P_2}$ (blue), (d) synchronization $S_{q}$, (e) entanglement $E_D$, with respect to time (in the units of $\tau=1/\omega_{m1}$). The parameters chosen are $ \omega_{m1}=-\Delta=1$, $\omega_{m2}=1.005$, $\bar n_{m}=0.5$ , $g_{1}=5\times10^{-5}$, $g_{2}=g_{1}\times10^{-2}$, $g_{3}=10^{-6}$, $\gamma_{mj}=0.009$, $\kappa=0.1$, $E=250$, $\eta_{D}=4$ and $\Omega_{D}=1$. All frequencies are normalized with respect to $\omega_{m1}$. (f) Evolution of time-averaged values of synchronization $\bar S_{q}$ (blue) and entanglement $\bar E_{D}$ (red) with respect to the frequency difference $\delta_{m}=\omega_{m2}-\omega_{m1}$ of the mechanical oscillators. [all the other parameters are the same as in (a)-(e) above].}
\label{g20}
\end{figure}

The initial states of the oscillators can be best approximated as Gaussian states with the peak at the mean position of the respective oscillators, along with minimum uncertainties in position quadratures. The fluctuation dynamics of the system is governed by a set of linearized equations, and this ensures that the evolved states also remain Gaussian. Since the Gaussian state can be fully characterized by its covariance matrix $C(t)$, 
we use it to calculate the correlation between the quantum fluctuations of the system variables and to compute  all the relevant measures, namely, $S_{q}$ and $E_{D}$. The  temporal dynamics of $C(t)$ is governed by the following linear ordinary differential equation:
\begin{eqnarray}
\dot{C}(t)=B(t) C(t)+B(t) C(t)^{ T }+\zeta\;,
\end{eqnarray}
where the elements of $C$ are given by $C_{ij}= \left[\left\langle M_i(\infty)M_j(\infty)+M_j(\infty)M_i(\infty)\right\rangle \right]/2 $ and the  matrix $\zeta$, as below, describes the diffusion of the system:
\begin{eqnarray}
       \zeta &=& {\rm diag}\big[
        0,(2 \bar n_{m1}+1)\gamma_{m1} ,
       0, (2 \bar n_{m2}+1)\gamma_{m2}, \kappa , \kappa \big]\;.
        \label{dmat}
\end{eqnarray}
In the matrix $C$, the diagonal (off-diagonal) element represents the variance of the respective mode (covariance of two modes). 

The QS quantifier  $S_{q}(t)$ can then be expressed as
\begin{eqnarray}
S_{q}(t)  = \left\{\frac{1}{2}\left[C_{11}(t)+C_{33}(t)-C_{13}(t)-C_{31}(t)+C_{22}(t)+C_{44}(t)-C_{24}(t)-C_{42}(t)\right]\right\}^{-1}\;,
\end{eqnarray}
and entanglement marker $E_{D}(t)$  as
\begin{eqnarray}
E_{D}(t)  = \frac{1}{4}\left[C_{11}(t)+C_{33}(t)-C_{13}(t)-C_{31}(t)\right]\times \left[C_{22}(t)+C_{44}(t)+C_{24}(t)+C_{42}(t)\right]\;.
\end{eqnarray}

\subsection{Numerical results} \label{numer1}
We show in Figs. \ref{g20}, the time-evolution of the quadratures of the oscillators, as obtained by simultaneously solving Eqs. (\ref{f1}). In Fig. \ref{pq}, we show how the evolution of $Q_{1}\rightleftharpoons P_{1}$ and $Q_{2}\rightleftharpoons P_{2}$ of the two oscillators tend to an asymptotic periodic orbit and the two limit cycles tend to be consistent with each other. This becomes apparent from the Figs. \ref{q1q2} and \ref{p1p2}, as the mean positions $Q_{1}$ and $Q_{2}$ and also the mean linear momenta $P_{1}$ and $P_{2}$ are found to oscillate exactly in phase at the long times. This refers to the complete classical synchronization between mechanical oscillators. 

With the classical limit cycle as a pre-condition, we next analyze how the $S_q$ and $E_D$ vary with time. From the Fig. (\ref{Sq}), we can see that $S_q$ reaches a stable value close to unity that refers to complete QS. More importantly, we find that this is associated with entanglement. From the Fig. (\ref{ED}), we can see that $E_D$ becomes less than 0.25. This shows that entanglement criterion (\ref{entcrit}) is violated by the state of the two oscillators. By comparing Figs. (\ref{ED}) and (\ref{Sq}), it can be observed that the generation of entanglement and synchronization between the oscillators occur at a similar time scale. This indicates a possible relation between the onset of QS and the generation of entanglement between the oscillators. In fact, the long-time  behavior, as displayed in Figs. (\ref{Sq}) and (\ref{ED}), clearly shows that $S_q$ and $E_D$ oscillate in similar way at a particular frequency.  

We further show in Fig. (\ref{delta}), how the QS and entanglement behave with increase in frequency difference $\delta_{m}=\omega_{m2}-\omega_{m1}$. The time-averaged values of the respective markers $S_q$ and $E_D$ for QS and the entanglement deteriorates with rise in frequency difference $\delta_{m}$. One can observe that the quantum complete synchronization (i.e., $\bar{S}_{q}=1$) and the maximal violation of entanglement inseparability (i.e., $\bar{E}_D-1/4$ is minimum) occur around the resonance condition $\delta_{m}=0$, i.e.,  for a pair of identical mechanical oscillators (i.e., $\omega_{m1}=\omega_{m2}$). Interestingly, they stay robust (i.e., $S_q$ remains less than 1 and $E_D$ less than 0.25) and therefore maintain both QS and entanglement, against the asymmetry in the oscillator frequencies, as large as $\delta_m \sim 3.2$. As $\delta_m\gtrsim 3.2$, the oscillators no longer remain entanglement (as $E_D$ exceeds 0.25) and quantum-synchronized (as $S_q$ exceeds 1) and both these features vanish together. This further strengthens our conjecture that QS and entanglement must originate from the same type of quantum correlation.

\section{Analytic solution of fluctuations}\label{analysis}
In order to have a deeper insight into entanglement and synchronization behaviour, we next obtain an analytical solution of the mean square fluctuations in the relative displacement $ q_{-}$, the total momentum $p_{+}$ and relative momentum $ p_{-}$. We start with the Eqs. (\ref{p1}), that can be rewritten for these variables as 
\begin{eqnarray}
\frac{d q_{-}}{d t} &=&\omega_{m} p_{-}\;,\nonumber\\
\frac{d q_{+}}{d t} &=&\omega_{m} p_{+}\;,\nonumber\\
\frac{d p_{-}}{d t}&=&-\omega_{m} q_{-}-2 g_{2} a^{\dagger} a q_{-}-g_{3} a^{\dagger} a q_{-}-\gamma_{m} p_{-}+\frac{1}{\sqrt{2}}\left(\xi_{1}-\xi_{2}\right)\;,\nonumber\\
\frac{d p_{+}}{d t}&=&-\omega_{m} q_{+}+\sqrt{2}g_{1} a^{\dagger} a-2 g_{2} a^{\dagger} a q_{+}+g_{3} a^{\dagger} a q_{+}-\gamma_{m} p_{+}+\frac{1}{\sqrt{2}}\left(\xi_{1}+\xi_{2}\right)\;,\nonumber\\
\frac{d a }{d t}&=&-(\kappa+\iota \Delta) a+\iota\left[ \sqrt{2}g_{1} q_{+}- g_{2}(q_{1}^{2}+q_{2}^{2})+ g_{3} q_{1}q_{2}\right]a
+E\left[1+\eta_{D} \cos\left(\Omega_{D}t\right)\right]+\sqrt{2 \kappa} a_{i n}\;, 
\label{eq:p2}
\end{eqnarray}
where we have assumed $\omega_{m1}=\omega_{m2}=\omega_{m}$, $g_1^{(j)} = g_1$ and $g_2^{(j)} =g_2$ for all $j$, for simplicity. 

\subsection{Asymptotic solutions of first moments}\label{meansec}
We next expand the operators $q_{\pm}$ and $p_{\pm}$ as a sum of their mean values and fluctuations as follows: $q_{\pm}=Q_{\pm}+\delta q_{\pm}$, $p_{\pm}=P_{\pm}+\delta p_{\pm}$. The Langevin equations (\ref{eq:p2}) can then be split into two sets:  one for the mean values, as listed below, and the other one for the fluctuation operators, to be discussed later.
\begin{eqnarray}
\frac{d Q_{-}}{d t} &=&\omega_{m} P_{-}\;,\nonumber\\
\frac{d Q_{+}}{d t} &=&\omega_{m} P_{+}\;,\nonumber\\
\frac{d P_{-}}{d t}&=&-\omega_{m} Q_{-}-2 g_{2} |A|^{2} Q_{-}-g_{3} |A|^{2} Q_{-}-\gamma_{m} P_{-}\;,\nonumber\\
\frac{d P_{+}}{d t}&=&-\omega_{m} Q_{+}+\sqrt{2}g_{1} |A|^{2} -2 g_{2} |A|^{2}  Q_{+}+g_{3} |A|^{2} Q_{+}-\gamma_{m} P_{+}\;,\nonumber\\
\frac{d A }{d t}&=&-(\kappa+\iota \Delta) A+\iota\left[ \sqrt{2}g_{1} Q_{+}- g_{2}(Q_{1}^{2}+Q_{2}^{2})+ g_{3} Q_{1}Q_{2}\right]A
+E\left[1+\eta_{D} \cos\left(\Omega_{D}t\right)\right]\;.
\label{eq:P2}
\end{eqnarray}

Since the cavity is driven by a modulating field $E(t)=E[1+\eta_{D}\cos\left(\Omega_{D}t\right)]=E_{0}+E_{1}e^{-i\Omega_{D}t}+E_{-1}e^{i\Omega_{D}t}$, the amplitudes of the cavity mode and the mechanical modes  would also follow the dynamics of this field at a long-time limit, according to the Floquet theorem, i.e., $\lim_{t\rightarrow\infty} A(t)=A(t+\tau)$, $\lim_{t\rightarrow\infty} Q_{1,2}(t)=Q_{1,2}(t+\tau)$ and $\lim_{t\rightarrow\infty} P_{1,2}(t)=P_{1,2}(t+\tau)$ \cite{mari2009gently}. We therefore redefine these classical amplitudes in terms of Fourier components to the first harmonic of the modulating frequency $\Omega_D$, as follows.
\begin{eqnarray}
    A & = & A_{-1}e^{i\Omega_{D}t} + A_{0} + A_{1}e^{-i\Omega_{D}t}\;,\nonumber\\
    Q_{-} & = & Q^{-}_{-1}e^{i\Omega_{D}t} + Q^{-}_{0} + Q^{-}_{1}e^{-i\Omega_{D}t}\;,\nonumber\\
    P_{-} & = & P^{-}_{-1}e^{i\Omega_{D}t} + P^{-}_{0} + P^{-}_{1}e^{-i\Omega_{D}t}\;,\nonumber\\
    Q_{+} & = & Q^{+}_{-1}e^{i\Omega_{D}t} + Q^{+}_{0} + Q^{+}_{1}e^{-i\Omega_{D}t}\;,\nonumber\\
    P_{+} & = & P^{+}_{-1}e^{i\Omega_{D}t} + P^{+}_{0} + P^{+}_{1}e^{-i\Omega_{D}t}\;.
    \label{eq:q2}
\end{eqnarray}

\subsubsection{Obtaining the Fourier coefficients of $Q_-$ and $P_-$}
To find the time-independent coefficients $Q_{-1,0,1}^{-}$, we first substitute $A$, $Q_{-}$, and $P_{-}$ from Eqs. (\ref{eq:q2}) into the Langevin equations (\ref{eq:P2}) for $Q_{-}$ and $P_{-}$. This leads us to the following set of three coupled nonlinear equations, when  $\Omega_D=\omega_{m}$:
\begin{eqnarray}
    \left(\omega_{m}+W\right)Q_{0}^{-}+W_0Q_{1}^{-}
    +W_0^*Q_{-1}^{-}&=&0\;,\nonumber\\
    W_0Q_{0}^{-}+W_1Q_{1}^{-}
    +\left(\iota\gamma_{m}+W\right)Q_{-1}^{-}&=&0\;,\nonumber\\
    W_0^*Q_{0}^{-}+\left(-\iota\gamma_{m}+W\right)Q_{1}^{-}
    +W_1^*Q_{-1}^{-}&=&0\;,
    \label{eq:Q6}
\end{eqnarray}
where \begin{eqnarray}
W_0&=&(2g_{2}+g_{3})(A_{0}A_{1}^{*}+A_{0}^{*}A_{-1})\;, \nonumber\\
W_1&=&(2g_{2}+g_{3})(A_{1}^{*}A_{-1})\;,\nonumber\\
W&=&(2g_{2}+g_{3})(|A_{0}|^{2}+|A_{1}|^{2}+|A_{-1}|^2)\;.
\end{eqnarray}
 The trivial solution of the above set of equations (\ref{eq:Q6}) is $Q_{-1,0,1}^{-}=0$, so that $Q_{-}$ vanishes. From the equation of $P_-$, we also have $P_-(t)=\exp(-\gamma_m t) P_-(0)$, which vanishes as $t\rightarrow \infty$.  Both the results $Q_-=0, P_-=0$  are expected in a synchronized system.  This further implies that $Q_{1}=Q_{2}=Q_{+}/\sqrt{2}$. 
 
 \subsubsection{Obtaining the Fourier coefficients $A_{-1,0,1}$}
 Putting the  above results of $Q_{1,2}$ in the Langevin equation of $A$ in (\ref{eq:P2}), we find that this reduces to the following equation:
 \begin{eqnarray}
\frac{d A }{d t}&=&-(\kappa+\iota \Delta) A+\iota\left[ \sqrt{2}g_{1} Q_{+}- \left(\frac{2g_{2}-g_{3}}{2}\right) Q_{+}^{2}\right]A
+E\left[1+\eta_{D} \cos\left(\Omega_{D}t\right)\right]\;,
\label{G4}
 \end{eqnarray}
 which contains only $Q_+$.
Substituting $A$ and $Q_+$ from the Eqs. (\ref{eq:q2}) into (\ref{G4}) and separating the different Fourier components, we obtain the following coupled equations of the time-independent coefficients $A_{-1,0,1}$:
\begin{eqnarray}
  UA_{0}+U_{-1}A_{1}
 +U_1A_{-1}&=&E_{0}\;,\nonumber\\
 U_1A_{0}+\left[-\iota \omega_m+U\right]A_{1}
 +\left[\frac{\iota}{2}(2g_{2}-g_{3})(Q_{1}^{+})^2\right]A_{-1}&=&E_{1}\;,\nonumber\\
U_{-1}A_{0}+\left[\iota \omega_m+U\right]A_{-1}
+\left[\frac{\iota}{2}(2g_{2}-g_{3})(Q_{-1}^{+})^ 2\right]A_{1}&=&E_{-1}\;,
\end{eqnarray}
where 
\begin{eqnarray}
    U &=&\left(\kappa+\iota\Delta\right)-\iota\sqrt{2} g_{1}Q_{0}^{+}+\iota(2g_{2}-g_{3})\left(\frac{1}{2}Q_{0}^{+ 2}+Q_{1}^{+}Q_{-1}^{+}\right)\;,\nonumber\\
    U_1&=&-\iota\sqrt{2} g_{1}Q_{1}^{+}+\iota(2g_{2}-g_{3})\left(Q_{0}^{+}Q_{1}^{+}\right)\;,\nonumber\\
    U_{-1}&=&-\iota\sqrt{2} g_{1}Q_{-1}^{+}+\iota(2g_{2}-g_{3})\left(Q_{0}^{+}Q_{-1}^{+}\right)\;.
\end{eqnarray}
Here we consider the weak optomechanical coupling regime, i.e., $|g_{i}| \ll \omega_m, \kappa$ $(i\in 1,2,3)$, such that $U\approx (\kappa +\iota\Delta)$ and $U_{\pm 1}$ may be neglected. Therefore, in the long-time limit,  we have
\begin{eqnarray}
    A_{0}=\frac{E_{0}}{\kappa+\iota\Delta},\;
    A_{1}=\frac{E_{1}}{\kappa+\iota(\Delta-\omega_{m})},\;
    A_{-1}=\frac{E_{-1}}{\kappa+\iota(\Delta+\omega_{m})}\;.\; 
    \label{eq:Q4}
\end{eqnarray}

\subsubsection{Obtaining the Fourier coefficients of $Q_+$ and $P_+$}
So far, we have obtained the expressions of the Fourier components of $Q_-$ and $A$. To find out $Q_+$, we again substitute $A$, $Q_{+}$, and $P_{+}$ from Eqs. (\ref{eq:q2}) into the Langevin equation (\ref{eq:P2}) for $Q_{+}$ and $P_{+}$, that leads us to the following set of three coupled linear equations of $Q^+_{-1,0,1}$:  
\begin{eqnarray}
 \left(\omega_{m}+V\right)Q_{0}^{+}+V_0Q_{1}^{+}+V_0^*Q_{-1}^{+}
  &=&\sqrt{2}g_{1}(|A_{0}|^{2}+|A_{1}|^{2}+|A_{-1}|^2)\;,\nonumber\\
 V_0Q_{0}^{+}+V_1Q_{1}^{+}+\left(\iota\gamma_{m}+V\right)Q_{-1}^{+}
  &=&\sqrt{2}g_{1}(A_{0}A_{1}^{*}+A_{0}^{*}A_{-1})\;,\nonumber\\
 V_0^*Q_{0}^{+}+\left(-\iota\gamma_{m}+V\right)Q_{1}^{+}+V_1^*Q_{-1}^{+}
  &=&\sqrt{2}g_{1}(A_{0}A_{-1}^{*}+A_{0}^{*}A_{1})\;,
    \label{eq:Q5}   
\end{eqnarray}
where \begin{eqnarray}
V_0&=&(2g_{2}-g_{3})(A_{0}A_{1}^{*}+A_{0}^{*}A_{-1})\;, \nonumber\\
V_1&=&(2g_{2}-g_{3})(A_{1}^{*}A_{-1})\;,\nonumber\\
V&=&(2g_{2}-g_{3})(|A_{0}|^{2}+|A_{1}|^{2}+|A_{-1}|^2)\;.
\end{eqnarray}
Using Eq. (\ref{eq:Q4}), we can find the time-independent coefficients $Q^+_{-1,0,1}$ by algebraically solving the above equations. The expressions of $P_+$ can then be easily obtained using the equation for $Q_+$ in (\ref{eq:P2}).

\subsection{Frequency spectrum of fluctuations}
To investigate the QS and entanglement between mechanical oscillators, we need to calculate the mean square fluctuations in $q_-$ and $p_\pm$. The dynamical behaviour of these fluctuations can be described by the following Langevin equations, as obtained from Eqs. (\ref{eq:p2}) [see Sec. \ref{meansec}]:
\begin{eqnarray}
 \frac{d}{d t} \delta q_{-} &=& \omega_{m} \delta p_{-}\;,\nonumber\\
 \frac{d}{d t} \delta p_{-} &=&-\left[\omega_{m}+\left(2g_{2}+g_{3}\right)\left(|A_{0}|^{2}+|A_{1}|^{2}+|A_{-1}|^2\right)\right]\delta q_{-}-\gamma_{m} \delta p_{-}+\frac{\xi_{1}-\xi_{2}}{\sqrt{2}}\;,\nonumber\\
 \frac{d}{d t} \delta q_{+} &=& \omega_{m} \delta p_{+}\;,\nonumber\\
 \frac{d}{d t} \delta p_{+} &=&-F_{0}\delta q_{+}+F_{1}\delta a^{\dagger}+F_{2}\delta a-\gamma_{m} \delta p_{+}+\frac{\xi_{1}+\xi_{2}}{\sqrt{2}}\;,\nonumber\\
 \frac{d}{d t} \delta a &=&-\left(\kappa+\iota\Delta^{\prime}\right)\delta a +\iota F_{1}\delta q_{+}+\sqrt{2 \kappa} \delta a_{i n}\;,\nonumber\\
 \frac{d}{d t} \delta a^{\dagger} &=&-\left(\kappa-\iota\Delta^{\prime}\right)\delta a^{\dagger}-\iota F_{2}\delta q_{+}+\sqrt{2 \kappa} \delta a_{i n}^{\dagger}\;,
 \label{eq:f4}
 \end{eqnarray} 
 where
 \begin{eqnarray}
 F_{0} &=& \omega_{m}+(2g_{2}-g_{3})\left(|A_{0}|^{2}+|A_{1}|^{2}+|A_{-1}|^{2}\right)\;,\nonumber\\
 F_{1} &=& \sqrt{2}g_{1}A_{0}-(2g_{2}-g_{3})\left(Q_{0}^{+}A_{0}+Q_{1}^{+}A_{-1}+Q_{-1}^{+}A_{1}\right)\;,\nonumber\\
 F_{2} &=& \sqrt{2}g_{1}A_{0}^{*}-(2g_{2}-g_{3})\left(Q_{0}^{+}A_{0}^{*}+Q_{1}^{+}A_{1}^{*}+Q_{-1}^{+}A_{-1}^{*}\right)\;,\nonumber\\  
 \Delta^{\prime} &=& \Delta-\sqrt{2}g_{1}Q_{0}^{+}+\left(\frac{2g_{2}-g_{3}}{2}\right)\left(Q_{0}^{+2}+2Q_{1}^{+}Q_{-1}^{+}\right)\;.
 \label{eq:p0}
 \end{eqnarray}
 These equation explicitly depend upon the Fourier components of $A$ and $Q_+$.
Introducing the cavity field quadratures $\delta x=\frac{\delta a+\delta a^{\dagger}}{\sqrt{2}}$ and $\delta y=\frac{i(\delta a^{\dagger}-\delta a)}{\sqrt{2}}$, and the input noise quadratures $\delta x_{in}=\frac{\delta a_{in}+\delta a_{in}^{\dagger}}{\sqrt{2}}$ and $\delta y_{in}=\frac{i(\delta a_{in}^{\dagger}-\delta a_{in})}{\sqrt{2}}$, the Eqs. (\ref{eq:f4}) can be reduced to the following matrix form:
 \begin{eqnarray}
     \dot{f}(t) = F f(t) + G(t)\;,
     \label{eq:p1}
 \end{eqnarray}
 where $f(t)$ is the fluctuation vector and $G(t)$ is the noise vector, with the respective transposes given by, 
 \begin{eqnarray}
     f(t)^{T} = \left(\delta q_{+}, \delta p_{+}, \delta x, \delta y\right),
     G(t)^{T} = \left(0, \frac{\xi_{1}+\xi_{2}}{\sqrt{2}}, \sqrt{2\kappa}\delta x_{in}, \sqrt{2\kappa}\delta y_{in}\right)\;,
     \label{eq:fB}
 \end{eqnarray}
 and the matrix $F$ is given by
 \begin{eqnarray}
 F=
  \begin{pmatrix}
   0 & \omega_{m} & 0 & 0\\
   -F_{0} & -\gamma_{m} & \frac{F_{1}+F_{2}}{\sqrt{2}} & \frac{i(F_{2}-F_{1})}{\sqrt{2}}\\
   \frac{-i(F_{2}-F_{1})}{\sqrt{2}} & 0 & -\kappa & \Delta^{\prime}\\
   \frac{F_{1}+F_{2}}{\sqrt{2}} & 0 & -\Delta^{\prime} & -\kappa\\
  \end{pmatrix} \;.
  \label{eq:p3}
 \end{eqnarray}

Taking the Fourier transformation of each operator in Eqs. (\ref{eq:f4}) and solving them in frequency domain, the fluctuations spectrum of the $q_-$ and $p_-$ can be obtained as
 \begin{eqnarray}
  \delta q_{-}(\omega) &=& \frac{-\omega_{m}}{d(\omega)}\left(\frac{\xi_{1}(\omega)-\xi_{2}(\omega)}{\sqrt{2}}\right)\;,\nonumber\\
  \delta p_{-}(\omega) &=& \frac{\iota\omega}{d(\omega)}\left(\frac{\xi_{1}(\omega)-\xi_{2}(\omega)}{\sqrt{2}}\right)\;,
  \label{f6}
\end{eqnarray}
 where $d(\omega)=\omega^{2}+\iota \omega \gamma_{m}-\omega_{m}^{2}-\omega_{m}(2g_{2}+g_{3})\left(|A_{0}|^{2}+|A_{1}|^{2}+|A_{-1}|^{2}\right)$. Similarly, the  fluctuation in the total momentum has the following spectrum:
 \begin{eqnarray}
  \delta p_{+}(\omega) &=& \frac{\iota\omega}{D(\omega)}\left[\sqrt{2\kappa}\left\{F_{1}[\kappa+\iota(\Delta^{\prime}-\omega)]\delta a_{in}^{\dagger}(-\omega)+F_{2}[\kappa-\iota(\Delta^{\prime}+\omega)]\delta a_{in}(\omega)\right\}+[(\kappa-\iota\omega)^{2}+\Delta^{\prime 2}]\frac{\xi_{1}(\omega)+\xi_{2}(\omega)}{\sqrt{2}}\right]\;,\nonumber\\
  \label{f7}
 \end{eqnarray}
 where $$D(\omega)=2\Delta^{\prime}\omega_{m}F_{1}F_{2}+[\omega^{2}+\iota\omega\gamma_{m}-\omega_{m}^{2}-\omega_{m}(2g_{2}-g_{3})\left(|A_{0}|^{2}+|A_{1}|^{2}+|A_{-1}|^{2}\right)][(\kappa-\iota\omega)^{2}+\Delta^{\prime2}]\;.$$ 

We must emphasize here that the stability of such a system is essential to achieve any synchronization. If all the eigenvalues of the matrix $F$ have negative real parts, the stability can be ensured at long times. Using  the Routh-Hurwitz criterion \cite{dejesus1987routh}, the corresponding conditions can be derived as follows:
 \begin{eqnarray}
    \kappa\gamma_{m}\left[\left(\Delta^{\prime 2}+\kappa^{2}\right)^{2}+\left(F_{0}\omega_{m}+\gamma_{m}\kappa\right)^{2}+2\gamma_{m}\kappa\left(\kappa^{2}+\Delta^{\prime 2}\right)\right.&&\nonumber\\
    \left.+2F_{0}\omega_{m}\left(\kappa^{2}-\Delta^{\prime 2}\right)+\gamma_{m}^{2}\Delta^{\prime 2}\right]
  +F_{1}F_{2}\Delta^{\prime}\omega_{m}\left(\gamma_{m}+2\kappa\right)^{2} &>&0\;,\nonumber\\
     F_{0}\omega_{m}\left(\kappa^{2}+\Delta^{\prime 2}\right)-2F_{1}F_{2}\omega_{m}\Delta^{\prime} &>&0\;.
     \label{eq:p4}
 \end{eqnarray}
We assume that the system satisfies these stability conditions.

 \subsubsection{Derivation of mean square fluctuations}
 The mean square fluctuation of an operator $O(t)$ is determined by \cite{huang2009entangling}
 \begin{eqnarray}
  \langle\delta O(t)^{2}\rangle &=& \frac{1}{4\pi^{2}} \iint_{-\infty}^{\infty}d\omega d\Omega e^{-\iota(\omega+\Omega)t}\langle\delta O(\omega)\delta O(\Omega)\rangle
 \label{f8}
 \end{eqnarray}
 where the $\langle\delta O(\omega)\delta O(\Omega)\rangle$ refers to two-frequency correlation function of the operator $O$. In the present case, we use Eq. (\ref{f6}) and (\ref{f7}) into Eq. (\ref{f8}) to obtain the following forms of mean square fluctuations:
 \begin{eqnarray}
  \langle\delta q_{-}(t)^{2}\rangle &=& \frac{1}{2\pi} \int_{-\infty}^{\infty} [\omega_{m}^{2}\mu]d\omega\;,\label{f09}\\
  \langle\delta p_{-}(t)^{2}\rangle &=& \frac{1}{2\pi} \int_{-\infty}^{\infty} [\omega^{2}\mu]d\omega\;,
  \label{f10}\\
  \langle\delta p_{+}(t)^{2}\rangle &=& \frac{1}{2\pi} \int_{-\infty}^{\infty} [\omega^{2}\nu]d\omega\;,
  \label{f11}
 \end{eqnarray}
 where 
 \begin{eqnarray}\mu &=&\frac{\gamma_{m}(2\bar n_{m}+1)}{d(\omega)d(-\omega)}\;,\nonumber\\
 \nu&=&\frac{1}{D(\omega)D(-\omega)}\left\{2\kappa F_{1}F_{2} [\kappa^{2}+(\Delta^{\prime}+\omega)^{2}]+\gamma_{m}(2\bar n_{m}+1)[(\Delta^{\prime2}+\kappa^{2}-\omega^{2})^{2}+4\kappa^{2}\omega^{2}]\right\}\;.
 \end{eqnarray} 
 
 In the above derivation of $\mu$ and $\nu$, we have used  the following non-vanishing frequency-domain correlation functions of the input noise operators, as obtained  by Fourier transformation of the Eqs. (\ref{G1}) and (\ref{g2}):
 \begin{eqnarray}
 \langle\delta a_{in}(\omega)\delta a_{in}^{\dagger}(-\Omega)\rangle & =&2\pi\delta(\omega+\Omega)\;,\nonumber\\
   \left\langle\xi_{j}(\omega) \xi_{k}\left(\Omega)\right)\right\rangle &=& 2\pi\delta_{jk}\gamma_{m}\left( 2\bar n_{ m }+1\right) \delta\left(\omega+\Omega\right)\;.
   \label{f9}
 \end{eqnarray}
 Here we have assumed $\bar{n}_{mj}=\bar{n}_m$ for all $j$, for simplicity.

 According to Wiener-Khintchine theorem \cite{mandel}, the functions $\omega_m^2\mu$, $\omega^2\mu$ and $\omega^2\nu$ represent the spectral density of the random fluctuation of $q_-$, $p_-$, and $p_+$, respectively. We find that they are even and converging functions of $\omega$, and therefore lead to finite values, when integrated over frequencies.
 \begin{figure}
        \subfloat[\label{Pp}]{%
		\includegraphics[height=4.5cm,width=.4\linewidth]{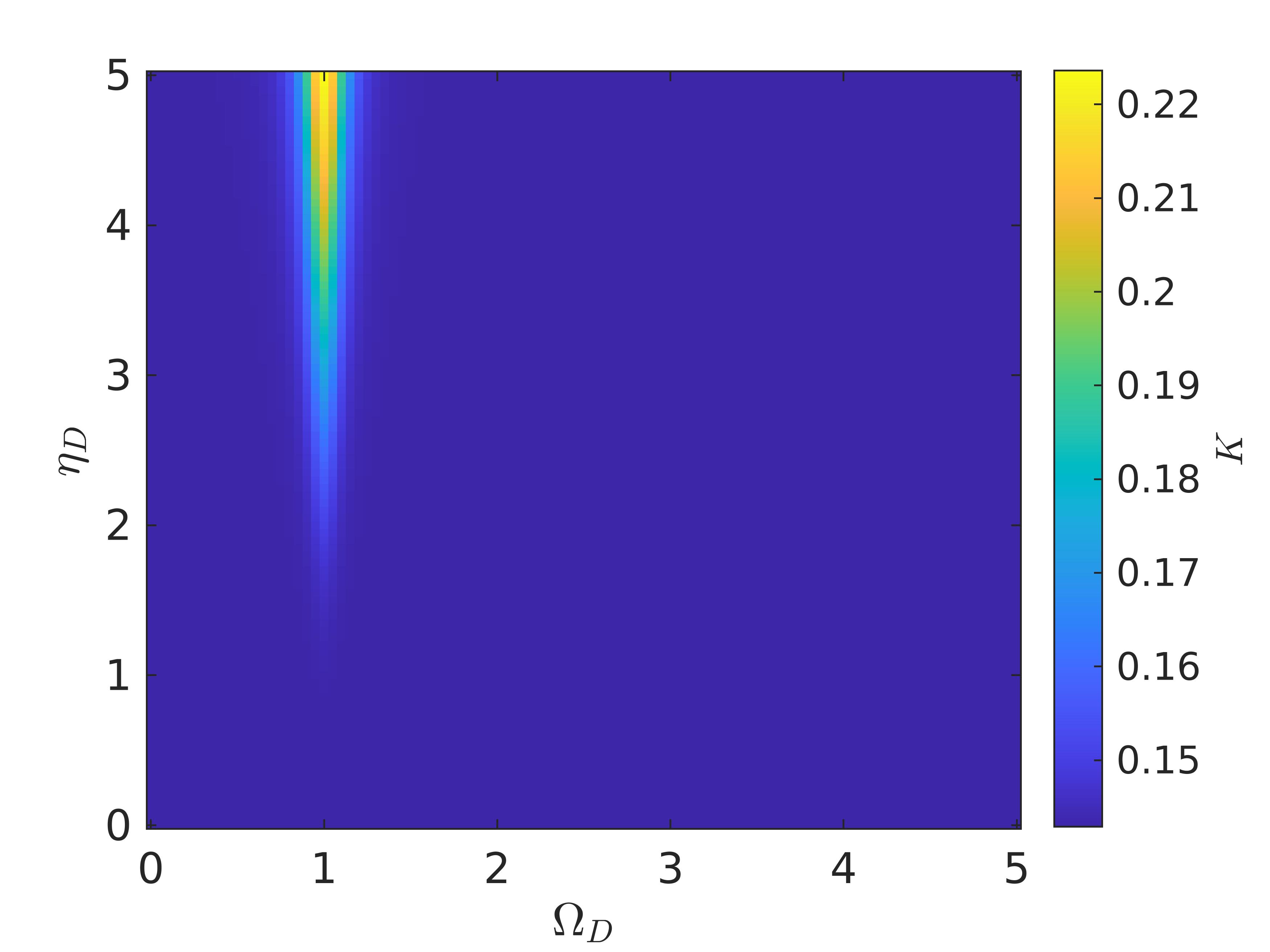}%
	}
	\subfloat[\label{SqD}]{%
		\includegraphics[height=4.5cm,width=.4\linewidth]{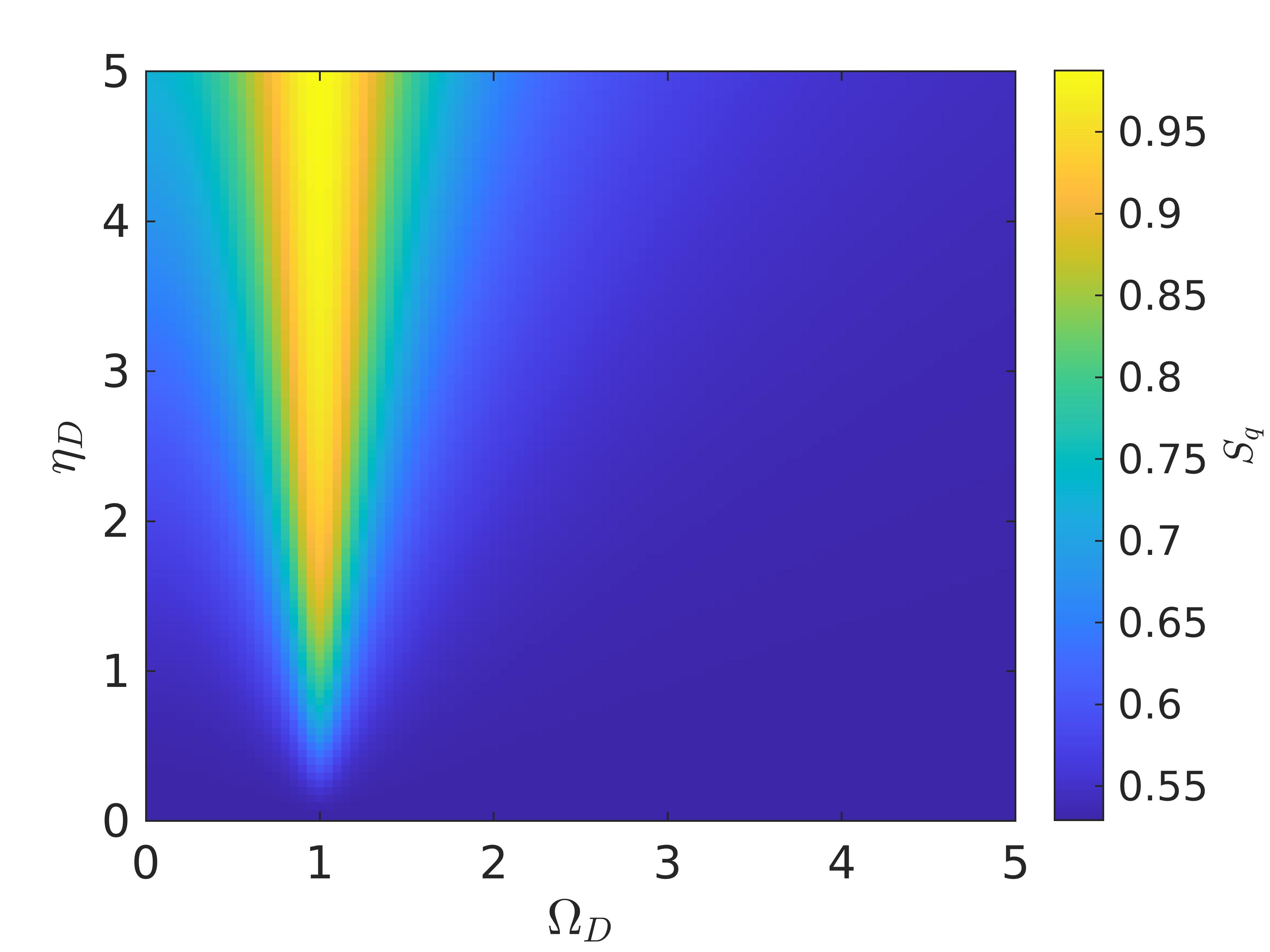}%
	}\\   
	\subfloat[\label{EdD}]{%
		\includegraphics[height=4.5cm,width=.4\linewidth]{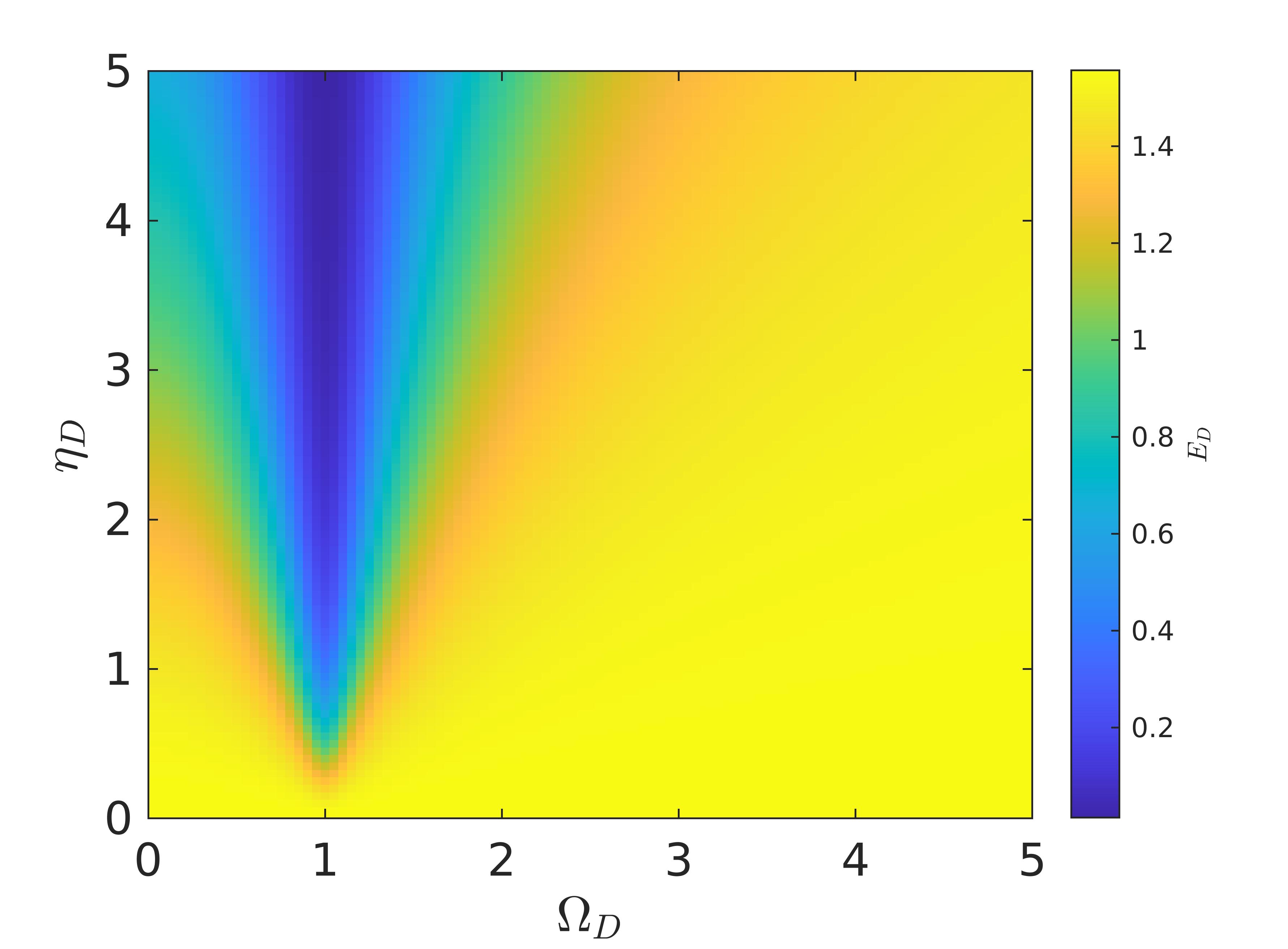}%
	}
	\hfill
	\caption{ Variation of (a) $K$ from Eq. (\ref{kcond}), (b) QS $S_{q}$, (c) entanglement $E_{D}$ between the mechanical oscillators as a function of modulation amplitude $\eta_{D}$ and modulation frequency $\Omega_{D}$ with $E=250$ and $\bar n_{m}=0.5$. The other parameters are the same as in Fig. \ref{g20}.}
	\label{wtemp}
\end{figure}
\subsection{Generic relation between entanglement and QS}\label{genericsec}
As seen in Eq. (\ref{SQ}),  $S_q$ always remains less than unity according to the uncertainty principle, and therefore, we find a lower limit for $\langle\delta q_{-}(t)^2\rangle$,  as given by
 \begin{equation}
     \langle\delta q_{-}(t)^2\rangle \ge 1-\langle\delta p_{-}(t)^{2}\rangle\;.
 \end{equation}
that corresponds to a state with QS. In a similar way, we can find out from (\ref{entcrit}) that there is an upper limit, as well, for $\langle\delta q_{-}(t)^2\rangle$, to achieve entanglement. This is given by
  \begin{equation}
     \langle\delta q_{-}(t)^2\rangle < \frac{1}{4\langle\delta p_{+}(t)^{2}\rangle}\;.\\
 \end{equation}
 Therefore, to make the entanglement and QS, we must have 
 \begin{equation}
 \frac{1}{4\langle\delta p_{+}(t)^{2}\rangle} > 1-\langle\delta p_{-}(t)^{2}\rangle\;,
 \end{equation}
 which can be simplified to the following analytical condition:
 \begin{equation}
   K=\frac{1}{4\langle\delta p_{+}(t)^{2}\rangle} + \langle\delta p_{-}(t)^{2}\rangle -1 > 0 \;.
   \label{kcond}
 \end{equation}
 When $K=0$, the two limits of $\langle\delta q_{-}(t)^2\rangle$ become equal, and it suggests a critical set of parameters, that are  required for simultaneous occurrence of entanglement and QS.

 We can finally express the QS $S_q$ in terms of $E_D$ as
 \begin{equation}
     S_q=\frac{\langle\delta p_{+}(t)^{2}\rangle}{E_D+\langle\delta p_{-}(t)^{2}\rangle\langle\delta p_{+}(t)^{2}\rangle}\;.
    \label{sqlimit}
 \end{equation}
 From the above expression, it is clear that when $E_D$ is less (a case when both $\langle\delta q_{-}(t)^{2}\rangle$ and $\langle\delta p_{+}(t)^{2}\rangle $ are less), the above expression of $S_q$ tends to $\langle\delta p_{-}(t)^{2}\rangle^{-1}$. Thus for smaller $\langle\delta p_{-}(t)^{2}\rangle$, $S_q$ increases. For sufficiently smaller $E_D$ (e.g., when it is less than 1/4), $S_q$ can increase towards unity [see Eq. (\ref{sqlimit}) above]. This means when the two oscillators become entangled, they can tend to near complete synchronization. On the other hand, for larger $E_D$ (e.g., for example, when it much larger than 1/4), the value $S_q$ becomes much less than unity. This suggests that in absence of entanglement, the degree of QS reduces.  Our analysis therefore suggests that entanglement becomes a sufficient condition for QS. In the next Subsection, we will explore this aspect more critically in our numerical studies. 

 \begin{figure}
        \subfloat[\label{sqE}]{%
		\includegraphics[height=4.5cm,width=.4\linewidth]{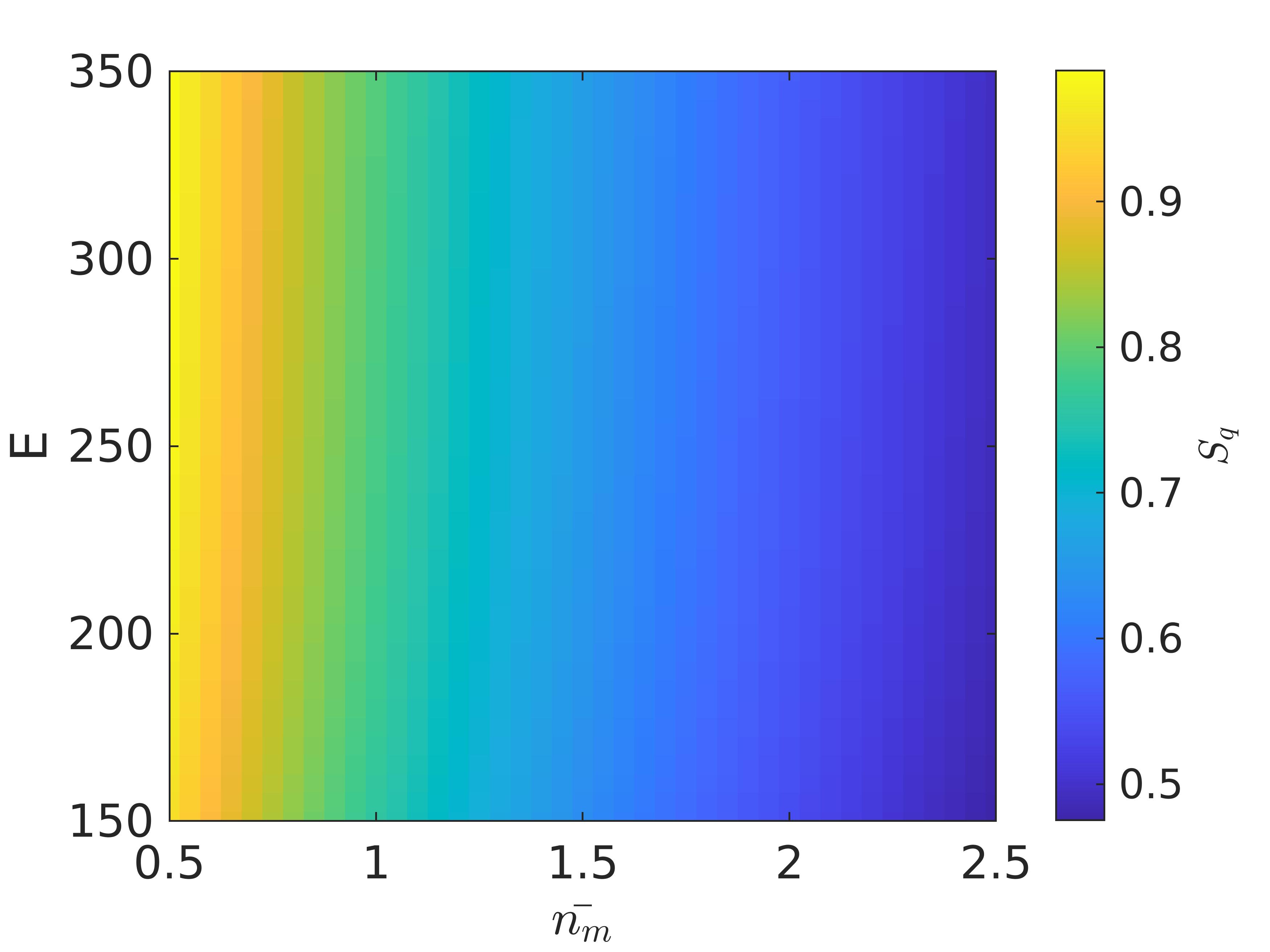}%
	}
	\subfloat[\label{edE}]{%
		\includegraphics[height=4.5cm,width=.4\linewidth]{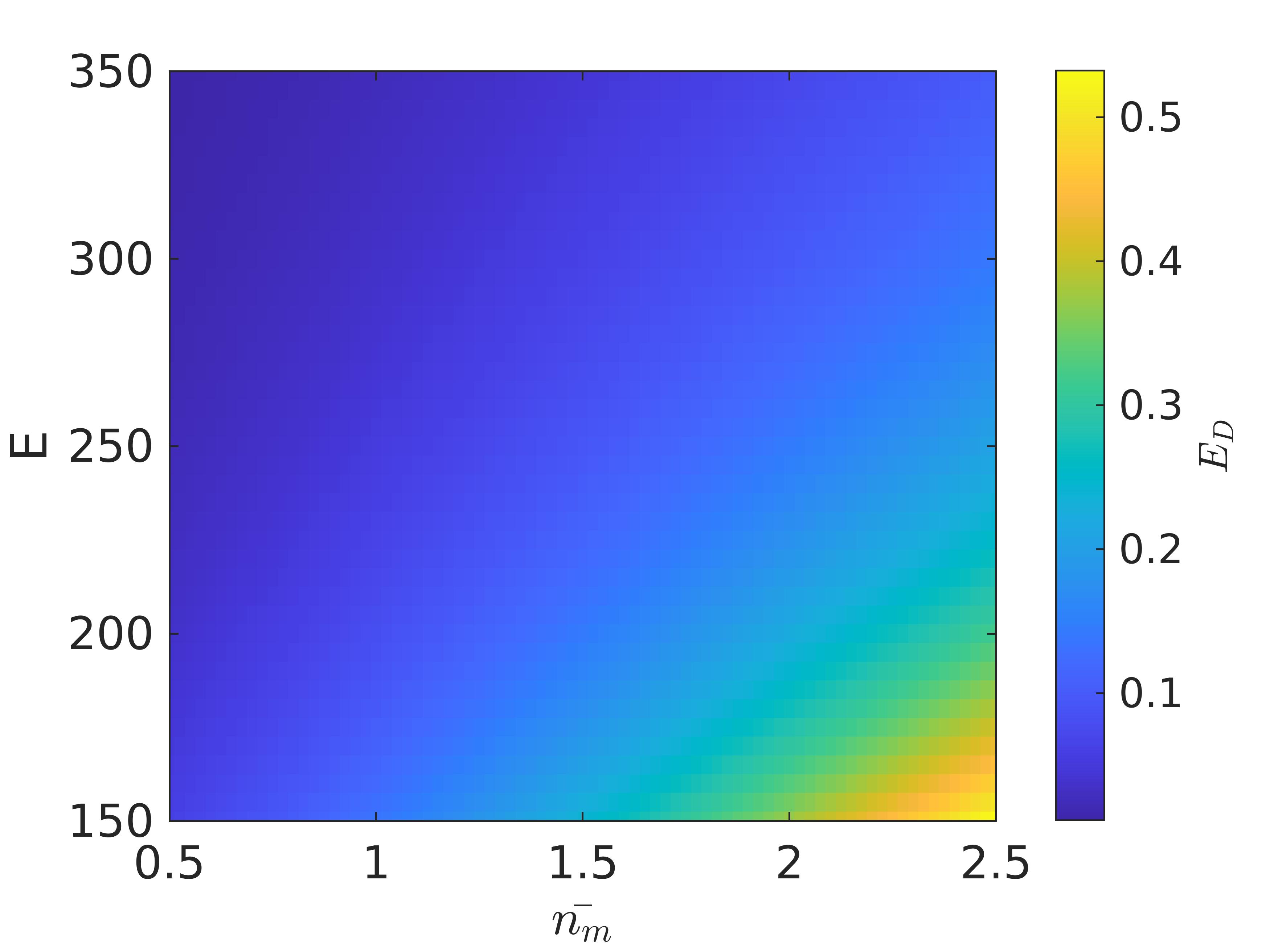}%
	}\\
	\hfill
	\caption{  Variation of (a) QS $S_{q}$ and (b) entanglement $E_{D}$ between the mechanical oscillators as a function of the driving field intensity $E$ and the average number of thermal phonons $\bar n_{m}$ with $\eta_{D}=4$ and $\Omega_{D}=1$.  The other parameters are the same as in Fig. \ref{g20}.}
	\label{wtemp1}
\end{figure}
 \subsection{Numerical results}
As noted in Secs. \ref{numer1} and \ref{genericsec}, there exists a correlation between QS and entanglement.  To test this conjectiure,  we present further numerical results based on the analytical results in Sec. \ref{analysis}. We choose all the parameters such that they maintain the stability in the system, following Eqs. (\ref{eq:p4}). We first show in Figs. (\ref{SqD}) and (\ref{EdD}), the variation of the QS $S_{q}$ and entanglement $E_{D}$, respectively, as a function of modulation amplitude $\eta_{D}$ and modulation frequency $\Omega_{D}$. It is observed that both the QS and the entanglement trace the classic 'Arnold tongue', yielding a range of values of $\eta_{D}$ and the corresponding values of $\Omega_{D}$ for which synchronization and entanglement are simultaneously achievable. We further find that the tongue is symmetric about $\Omega_D=\omega_m=1$. The  synchronization $S_q$ tends to one and the entanglement $E_D$  decreases further below the upper limit of 0.25, as $\eta_D$ increases, i.e., as the modulation strength increases. We note that similar entanglement tongue was reported for two coupled van der Pol oscillators in \cite{lee2014}. However, this was shown associated only with classical phase synchronization. 

We show in Fig. (\ref{Pp}) the variation of $K$ [Eq. (\ref{kcond})] as a function of $\eta_{D}$ and $\Omega_{D}$. This plot also exhibits the classic 'Arnold tongue' around modulation frequency $\Omega_{D}=1$.  For larger $\eta_{D}$ at $\Omega_{D}=1$, the term $K$ increases further. By comparing with the Figs. \ref{SqD} and \ref{EdD}, we can conclude that with increase in $K$  the system approaches to nearly complete QS (i.e., $S_q\lesssim 1$) and larger violation of the inseparability inequality (\ref{entcrit}) (i.e., $E_D \ll 0.25$). It is also interesting to see that $K$ remains positive for a large range of $\eta_D$ and $\Omega_D$. Thus, both the QS (albeit partial) and entanglement stay robust over a large variation of modulation parameters. 

To get further insight into the correlated behaviour of synchronization and entanglement, we plot them with respect to the driving field intensity $E$ and average number of phonons $\bar n_{m}$, respectively, in the Figs. (\ref{sqE}) and (\ref{edE}), for a fixed modulation amplitude $\eta_{D}=4$ and modulation frequency $\Omega_{D}=1$. It can be seen, for higher $E$ and at lower temperatures of the bath ($\bar n_{m}=0.5$), the oscillators are nearly complete quantum synchronized, along with maximal violation of inseparability inequality (\ref{entcrit}). With increase in thermal excitation, the synchronization deteriorates and the entanglement decreases too. However, the synchronization stays more robust to temperature as compared to entanglement. This indicates that though some residual synchronization may still exist without entanglement, the entanglement is always associated with the synchronization and therefore becomes the sufficient condition for the latter. In fact, this can also be intuitively understood from the definition (\ref{SQ}) of QS, which always remains less than unity due to uncertainty principle, though the quantum state of the system may not exhibit entanglement [i.e., may not violate the Eq. (\ref{entcrit})]. It must be borne in mind that the entanglement would indicate the existence of QS, only in the presence of limit cycle of mean values of the joint quadratures.
  \section{CONCLUSION}
  In conclusion, we have explored the interconnection between QS and entanglement between two mechanical oscillators in an optomechanical system. In our model, both mechanical oscillators are coupled with the same cavity mode via linear and quadratic dependence on their displacement from their respective equilibrium positions. An indirect always-on coupling, proportional to $g_3a^\dag a$ also arises between the oscillators, which is the key to generation of the synchronization between them. Since the same uncertainty relation sets the upper limit for QS $S_{q}$ and lower limit for entanglement marker $E_{D}$, so we expected that there could be a correlation between the two. In this regard,  we have first demonstrated the classical synchronization via limit cycle trajectories of the mean quadratures at long times, which is a precondition to achieve QS. Our numerical results show that the two coupled mechanical oscillators exhibit entanglement and QS, when the cavity is strongly amplitude-modulated. A nearly complete QS and entanglement between the coupled oscillators can be achieved for a large range of modulation parameters. This result leads us to a strong conjecture  that they arise from the same EPR-type correlation. We further show that the entanglement manifests itself as the sufficient condition for QS. One can characterize the QS as a function of entanglement, when defined in terms of variances of quadratures.  We have provided all the relevant analysis, supported by numerical results in this context. Our results open up a newer perspective to interpret the QS and entanglement on the same footing.
 \section*{References} 
\bibliography{References}
 \bibliographystyle{ieeetr}
\end{document}